\documentstyle[amssymb,floats,aps,prl,epsf]{revtex}
\epsfclipon 

\begin{document}

\twocolumn[\hsize\textwidth\columnwidth\hsize\csname 
@twocolumnfalse\endcsname 

\draft
\title{Mesoscopic Tunneling Magnetoresistance}
\author{Gonzalo Usaj and Harold U. Baranger}
\address{Department of Physics, Duke University, Box 90305, 
         Durham NC 27708-0305, USA}
\date{June 27, 2000; revised November 26, 2000}
\maketitle

\begin{abstract}
We study spin-dependent transport through
ferro\-mag\-net/nor\-mal-metal/ferro\-mag\-net double tunnel junctions in
the mesoscopic Coulomb blockade regime.  A general transport equation allows
us to calculate the conductance in the absence or presence of spin-orbit
interaction and for arbitrary orientation of the lead magnetizations. The
tunneling magnetoresistance (TMR), defined at the Coulomb blockade
conductance peaks, is calculated and its probability distribution presented. 
We show that mesoscopic fluctuations can lead to the optimal value of the TMR.
\end{abstract}
]
\bigskip
\narrowtext

Spin-polarized electron tunneling has become a very active area
of research, motivated by its potential role in new electronic
devices\cite{Prinz} as well as by the possibility of observing novel
effects\cite{Johnson,Ono,Fert,Ralph}. In ferro\-mag\-net/ferro\-mag\-net single
tunnel junctions (F-F, separated by an insula\-ting barrier), the subject
has been investigated for some time\cite{Julliere}, their
main feature is that the resistance depends on the relative orientation of
the magnetization in the ferromagnets. This is characterized by the tunneling
magnetoresistance (TMR), defined as the relative change of the resistance
when the magnetizations of the ferromagnets rotate from being parallel
to anti-parallel. 

A richer variety of effects are observed in
ferro\-mag\-net/nor\-mal-metal/ferro\-mag\-net double tunnel junctions
(F-N-F)  as well as in completely ferromagnetic double junctions (F-F-F),
due to the interplay between charge and spin. These systems consist of two
ferromagnetic leads separated by a metallic grain.  If the capacitance of
the grain is such that the thermal energy $k_{B}T$ is smaller than the
charging energy $E_{C}$, then the Coulomb blockade (CB) of electron
tunneling occurs. Specifically, the conductance shows pronounced peaks as
a function of an external gate controlling the electrostatic potential on
the grain.  How this charging effect modifies the TMR has been the subject
of recent experimental~\cite{Ono,Fert} and theoretical
\cite{Takahashi,Wang,Barnas,Brataas2,Inamura} work. In F-F-F
systems, the TMR is enhanced in the Coulomb blockade
regime by higher-order tunneling processes such as co-tunneling
\cite{Ono,Takahashi,Wang,Barnas} while in F-N-F systems the TMR is
reduced~\cite{Brataas2}. The effects of quantum interference within the Coulomb blockade has not been
considered in these systems as of yet.

On the other hand, the quantum Coulomb blockade has been extensively
investigated in the absence of ferromagnets~\cite{Jalabert,CBrevs}.  Quantum
interference in these structures necessarily entails mesoscopic fluctuations
as the detailed shape of the grains cannot be controlled.  Such effects
become important for sufficiently low temperature or, at room temperature,
for sufficiently small systems. In such a quantum regime ($k_{B}T \!\ll\!
\Delta  $), the CB conductance peak height, $G_{\rm peak}$, fluctuates
strongly as a function of the gate voltage~\cite{Jalabert,CBrevs}.

Here we study a problem at the intersection of these two fields, quantum
mesoscopics and spintronics.  Specifically, we study quantum effects in the
TMR at the CB peaks---the \emph{mesoscopic tunneling magnetoresistance}---in
F-N-F systems.  In the case of zero spin-orbit tunneling, we show that there
are CB peaks with the {\it maximum} possible value of TMR; in contrast to the
classical case, no special tuning is required.  For strong spin-orbit
coupling, the TMR can be {\it large} despite the SO coupling.  As for $G_{\rm
peak}$, the mesoscopic TMR is characterized by a probability distribution
rather than by a single number, and we obtain this distribution in both cases
above.

Since $G_{\rm peak}$ depends on the properties of a \emph{single} wave
function, one can legitimately wonder if it is possible to probe its spin
structure (a very active subject~\cite{Ralph,Baranger,SP-STM}) through the
conductance in non-collinear configurations.  For such a configuration, we
calculate $G_{\rm peak}$ as a function of the angle between the
magnetizations in the two leads. In this case as well as for strong SO, the
usual description in terms of a rate-equation is not possible, and a more
general equation for the conductance is presented. Finally, we study the case
of broken spin degeneracy in the grain and point out that it is equivalent to
having a half-metallic~\cite{halfmetallic} lead.

Figure 1 shows a schematic of the system: a normal metal grain is
weakly coupled to two ferromagnetic reservoirs by tunnel junctions.  The
potential of the grain is controlled by gate voltage $V_{g}$.  We consider
$\Gamma \!\ll\! k_{B}T \!\ll\! \Delta \!\ll\! E_{C}$, where
$\Gamma$ is the total width of the resonant levels in the dot. In this
regime, only a single energy level contributes to the conductance. Because of
time-reversal symmetry, each level is doubly degenerate (Kramers doublet).
Within a Hartree-Fock treatment of the electron-electron interactions, the
transport is described in terms of the self-consistent single electron wave
functions;  for the resonant doublet, these are the spinors
$\Psi_{1}({\mathbf r}) \!=\! (\phi ({\mathbf r}),\chi ({\mathbf r}))^{\mathrm
T}$ and $\Psi_{2}({\mathbf r}) \!=\! (-\chi ^{\ast }({\mathbf r}),\phi ^{\ast
}({\mathbf r}))^{\mathrm T}$.  The mesoscopic tunneling magnetoresistance is 
defined as
\begin{equation}
\label{tmr}
\eta =\frac{G_{P}-G_{AP}}{G_{AP}}
\end{equation}
where $G_{P(AP)}$ is the conductance at the CB peak in the 
parallel (antiparallel) configuration.

\begin{figure}
\begin{center}
\leavevmode
\epsfxsize = 7.0cm 
%\hsize
\epsffile[109 384 505 568]{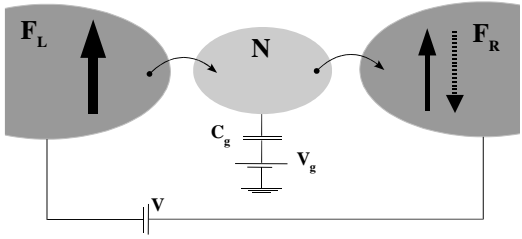}
%\epsfbox{levels.eps}
\end{center}
\caption{Schematic picture of the F-N-F double junction. A normal metal
grain is coupled to ferromagnetic leads by tunnel junctions and is small
enough for charging to be important. When the magnetization of one
lead is flipped, the conductance changes: the relative change with
respect to the parallel configuration defines the TMR.} 
\end{figure}

{\it Absence of Spin-Orbit---}
Consider the simplest case of no SO interaction and
collinear magnetizations. Then, the Kramers doublet corresponds to two-fold spin
degeneracy, $\chi({\mathbf r})\!\equiv\!0$, assuming the Zeeman energy
associated with any magnetic field present is much smaller than $k_{B}T$.
Linear-response theory yields for the conductance\cite{Beenakker}
\begin{equation}
\label{G}
G_{\mathrm peak}=\frac{e^{2}\ \lambda}{\hbar\ k_{B}T}\sum_{m=1,2}\frac{\,
\Gamma_{m}^{L}\, \Gamma_{m}^{R}}{\Gamma_{m}^{L}+\Gamma_{m}^{R}}
\end{equation}
where the sum is over the spin degenerate states and $\lambda$ is a numerical
factor\cite{Beenakker,Glazman,factor}. The partial width $\Gamma^q_{m}$, $q\!=\!L(R)$, due
to tunneling to the left (right) electrode is $\Gamma_{m}^q \!=\!
2\pi \, \sum_{\alpha}\rho_{\alpha}^q| V_{\alpha m}^q| ^{2}$ with $V_{\alpha m}^q $
the matrix element between $\left| m\right\rangle$ in the
grain and the channels $\alpha \!= \uparrow ,\downarrow$ in lead $q$, and
$\rho_{\alpha }^q$ the local density of states of those channels. For
point-contact leads,  $\Gamma_{1(2)}^q\!=\!|\phi_q| ^{2}\, t_{\uparrow
(\downarrow )}^q$  where $| \phi_q| ^{2}$ is the wave function of the resonant
level at the contact point ${\mathbf r}_q$ and $t_{\alpha}^q $ is a factor
which depends on  both $\rho_{\alpha}^{q}$ and the tunnel barrier. Notice we
assume that the tunneling through the barriers conserves spin. Letting $D_q$
($d_q$) denote $t^q_\alpha$ for the majority (minority) spins, we find
\begin{equation} \label{TMR}
\eta =\frac{(D_{L}-d_{L})(D_{R}-d_{R})| \phi_L| ^{2}| \phi_R| ^{2}}{(D_{L}| \phi_
L| ^{2}+D_{R}| \phi_R| ^{2})(d_{L}| \phi_L| ^{2}+d_{R}| \phi_R| ^{2})}.
\end{equation}

Since metallic grains are generally irregular in shape, one expects the
chaotic nature of the single particle classical dynamics to produce random
matrix statistics for the wave functions.  Following 
\cite{Jalabert}, we use RMT to describe the fluctuations of $| \phi_q |
^{2} $: the Gaussian Orthogonal Ensemble (GOE) for the Hamiltonian
implies the Porter-Thomas distribution, ${\cal P}(| \phi_q | ^{2})\propto |
\phi_q |^{-1}\, \exp (-| \phi_q | ^{2})$.  Assuming 
no correlation between $| \phi_{L}| ^{2}$ and $| \phi_{R}| ^{2}$ because
the contact points are far apart,
we obtain
\begin{equation}
\label{GOEgeneral}
{\cal P}_{\mathrm GOE}(\eta )=\frac{a}{\pi|\eta_{\rm m}^{}|}\frac{\Theta
(1-\eta/\eta_{\rm m}^{})}
    {\sqrt{1-\eta/\eta_{\rm
m}^{}}\sqrt{\eta/\eta_{\rm m}^{}}\ [1+\eta/\eta_{\rm m}^{} (a^{2}-1)]}
\end{equation}
 with 
\begin{eqnarray}
\nonumber
a & = & (\zeta ^{1/4}+\zeta ^{-1/4}) / (\delta^{1/4}+\delta ^{-1/4}) \;,\\
\zeta  & = & D_{R}d_{R}/D_{L}d_{L}\;, \qquad
\delta = D_{R}d_{L}/D_{L}d_{R}  \;,\\
\eta_{\rm m}^{} & = & 2P_{L}P_{R}/[1-P_{L}P_{R}+
   [(1-P^{2}_{L})(1-P^{2}_{R})]^{1/2} ] \;.
\nonumber
\end{eqnarray}
Here 
\begin{equation}
   P_q=\frac{D_q-d_q}{D_q+d_q}\;, \qquad q=L,R
\end{equation}
is the spin-polarization of the electrons coming from the leads\cite{Teresa}. Note that in the
case of symmetric barriers and the same ferromagnets---and only in this
case---${\cal P}_{\mathrm GOE}(\eta)$ 
factorizes into a term depending on the
polarizations $P_q$ and one depending on the properties of the grain.  

The maximum value of the TMR, $\eta_{\rm
m}^{}$, is also the upper limit for the TMR   when $k_{B}T\gg E_{C}$. In that
regime,  it is reached when $\zeta \!=\! 1$ (optimal asymmetry) while in our
case the condition is $|\phi_L |^2 / |\phi_R |^2 \!=\! \zeta$.  If the two
leads are made of different materials, the former condition requires that the
tunneling barriers be  expressly designed while in the mesoscopic regime there
are always  fluctuations that satisfy the latter condition. Indeed, note that
the  probability of the TMR being close to the maximum value is not small. In
this  case, then, {\it the mesoscopic fluctuations  lead to the optimal
value of the TMR}. 

{\it Broken Spin Degeneracy---} The spin degeneracy can be removed by
applying a magnetic field such that the Zeeman energy exceeds $k_B T$.  Then,
only one channel contributes to (\ref{G}):  the resonant level acts as an
ideal spin filter rendering the polarization of the left lead irrelevant. In
this way an effective {\em half-metallic}~\cite{halfmetallic} injector, an
object of considerable interest in the spintronics community, can be
constructed in conventional materials.  Proceeding as before, we find for 
the TMR
\begin{equation}
\eta = \frac{ \eta_{\rm m}^{}| \phi_L| ^{2} }
         {| \phi_L| ^{2}+| \phi_R| ^{2}D_{R}/D_{L} } \quad \mbox{with} \quad
       \eta_{\rm m}^{} = \frac{ 2P_{R} }{ 1-P_{R} } \;.
\end{equation}
Results for this case can be obtained from the spin degenerate results by
formally taking $d_{L}\!\rightarrow\!0$. This yields the distribution
Eq. (\ref{GOEgeneral}) and the mean value $\left\langle \eta
\right\rangle \!=\! \eta_{\rm m}^{}/(1+a)$, as before.  {\it Note that for
$D_{R} \!\ll\! D_{L}$, the \emph{average} TMR is much bigger than the
symmetric single tunnel-junction result}\cite{Julliere}:  $ \left\langle \eta
\right\rangle \simeq \eta_{\rm m}$ compared to $2P_{R}^{2}/(1-P_{R}^{2})$, a
factor $(1+1/P_{R})$ larger.
For ferromagnetic leads made of Co ($P \! \simeq\! 0.35$) this means an 
enhancement of $300\%$. The magnetic field here must,of course, be smaller 
than the coercive field of  the electrode (in the $AP$ configuration), making 
low temperatures advantageous in observing this effect 
(typically, $B\lesssim 500 \,{\mathrm G}$ and $T\!\lesssim\!70\,{\mathrm mK}$).
 
{\it Spin-Orbit Coupling---} SO coupling causes the direction of the spin of
an electron to rotate while in the grain, so that the natural axis
of quantization on the left is not the same as on the right.  Time-reversal
symmetry still applies but now neither component of the spinor $(\phi
({\mathbf r}),\chi ({\mathbf r}))^{\mathrm T}$ vanishes.  Since
the quantization axes in the leads are \emph{fixed} by the bulk
magnetization of the ferromagnets, the electrons tunnel into a superposition
of ``up'' and ``down'' states. This \emph{coherence} cannot be included in a
simple rate equation---a more general approach is
required.  Technically, the mixing of the spin channels means that
$\Gamma^L$ and $\Gamma^R$, which in the general case are matrices ${\mathbf
\Gamma}_{nm}^{q} \!=\! 2\pi \sum_{\alpha}\rho_{\alpha } ^q V_{\alpha n}^{q\ast
}V_{\alpha m}^q$, cannot be diagonalized in the same basis.  In the relevant $2
\times 2$ subspace ${\mathbf \Gamma}^{q}$ takes the form
\begin{equation}
{\mathbf \Gamma}^{q}=\left(\begin{array}{cc}
   \left|\phi_{q}\right|^{2}t_{\uparrow}^q + 
   \left|\chi_{q}\right|^{2}t_{\downarrow}^q & 
   \chi_{q}^{\ast}\phi_{q}^{\ast}(t_{\downarrow}^q-t_{\uparrow}^q) \\
   \chi_{q}\phi_{q}(t_{\downarrow}^q-t_{\uparrow}^q) & 
   \left|\chi_{q}\right|^{2} t_{\uparrow}^q + 
   \left|\phi_{q}\right|^{2}t_{\downarrow}^q \\
\end{array}\right)\, .  
\label{gama} 
\end{equation} 
Note that $\left[ {\mathbf \Gamma}^{L},{\mathbf \Gamma}^{R}\right]\neq 0$
only when \emph{both} leads are ferromagnetic \emph{and} SO coupling is
present.

From the Keldysh formalism, the current through an arbitrary 
system connected to two leads is\cite{Meir,Pastawski,Davies}
\begin{equation}
J\!=\!\frac{e}{2 h}\int\!{\mathrm d}\varepsilon\, {\mathrm
Tr}\left\{\left({\mathbf \Gamma}^{L}_{}\!-\!{\mathbf \Gamma}^{R}_{}\right){\mathrm i}{\mathbf
G}^{<}_{}\!+\!\left({\mathbf \Gamma}^{L}_{}f_{L}^{}\!-\!{\mathbf
\Gamma}^{R}_{}f_{R}^{}\right)\!{\mathbf A}\right\} 
\label{J} 
\end{equation}  
where 
${\mathbf A}\!=\! {\mathrm i}({\mathbf G}^{r}\!-\!{\mathbf G}^{a})$ 
is the spectral function, ${\mathbf G}^{r(a)}$ is the retarded (advanced)
many-body Green function of the full system, and $f_{L(R)}$ is the Fermi
distribution in the left (right) lead. 
It is convenient to express ${\mathbf G}^{<}$, the lesser Green function,  
in terms of the operators
${\mathbf I}_{}^{q} \!\equiv\! 
{\mathbf \Gamma}^{q}({\mathbf G}^{<}-{\mathrm i} f_{q}{\mathbf A})$:
${\mathbf G}^{<} \!=\! ({\mathbf \Gamma}^{L}+{\mathbf \Gamma}^{R})^{-1}
({\mathbf I}_{}^{L}+{\mathbf I}_{}^{R}+{\mathrm i}({\mathbf \Gamma}^{L}f_{L}
+{\mathbf \Gamma}^{R}f_{R}){\mathbf A})$. For elastic transport and within the
HF approximation, $({\mathbf I}_{}^{L}+{\mathbf I}_{}^{R})_{nn} \!=\! 0$, so
that  Eq. (\ref{J}) becomes  
\begin{equation} 
J=\frac{e}{h }\int \!\! {\mathrm d}\varepsilon\, {\mathrm Tr}\left\{{\mathbf
\Gamma}^{R}\left({\mathbf \Gamma}^{L}+{\mathbf \Gamma
}^{R}\right)^{-1}{\mathbf \Gamma}^{L}{\mathbf
A}\right\}\left(f_{L}-f_{R}\right) \,. 
\label{J2}
\end{equation} 
This equation simplifies for linear-response and weak coupling.
${\mathbf A}$ is then evaluated in equilibrium: it
depends on the equilibrium populations of the grain eigenstates
(which depend only on the energy) and on the
eigenfunctions of the isolated grain \cite{Meir}. Therefore, it can be
shown that within the resonant subspace ${\mathbf A}$ is proportional to the
identity. For the conductance, we finally get
 \begin{equation}
 G_{\mathrm
peak}=\frac{e^{2}\ \lambda'}{\hbar\ k_{B}T}{\mathrm Tr}\left\{{\mathbf
\Gamma}^{R}\left({\mathbf \Gamma}^{L}+{\mathbf \Gamma}^{R}\right)^{-1} {\mathbf
\Gamma}^{L}\right\} \label{G2}  
\end{equation}
where the numerical factor $\lambda'$ has the same origin as $\lambda$ in
(\ref{G}). Eq. (\ref{G2}) contains (\ref{G}) as a special limit; as far as we
know, it is presented in
this form for the first time.

Since $G_{\mathrm peak}$ can be calculated in any basis, we choose one where $\chi_L\!=\!0 $. 
Then, for symmetric leads and barriers ($P_{q} \!=\! P$, $\zeta\!=\!1$) the 
TMR from Eqs. (\ref{gama}) and (\ref{G2}) is
\begin{equation}
\eta = \frac{2\ P^2 \cos\theta_{\rm g}}{1+(1-P^2)\ \xi-P^2 \cos\theta_{\rm g}}
\label{GSE} 
\end{equation}
where $\theta_{\rm g}$ is the azimuthal angle of the spinor $\Psi_{1}({\mathbf r}_{R})$ and
\begin{equation}
\xi=\frac{|\Psi_{L}|^4+|\Psi_{R}|^4}{2\ |\Psi_{L}|^2|\Psi_{R}|^2}
\end{equation}
with $ |\Psi_{q}|\!=\!|\Psi_{1}({\mathbf r}_{q})| $. Note that $\eta\!=\!0$
if the spin at the right contact is perpendicular to the polarization in the
leads. For simplicity
we will assume that SO is strong and so describe the
statistics by the
 Gaussian Symplectic Ensemble (GSE). The intermediate case
is straightforward  numerically. Then, $\phi$ and $\chi$ are uncorrelated
complex random variables. Using their Gaussian distributions we find ${\cal
P}(\xi)$ and show that $ {\cal P}(\xi,\cos{\theta_{\mathrm g}})\!=\!{\cal
P}(\xi)/2 $. Therefore, the distribution of the TMR is
\begin{equation}
{\cal P}_{\mathrm GSE}(\eta)\!=\!\frac{1}{2\ \eta_{\mathrm m}^{3/2}(2+\eta)^2}
\times\!\left\{ \begin{array}{ll} 
{\cal A}(\eta), & 
0\!\le\!\eta\!\le\!\eta_{\mathrm m} \\ 
&
\\
{\cal A}\left(-\frac{\eta}{1+\eta}\right), &
-P^2\!\le\!\eta\!\le\!0  \\
\end{array}\right.
\end{equation}
where ${\cal A}(x)\!=\!\sqrt{\eta_{\mathrm m}-x}\
(6+2\eta_{\mathrm m}+x)$ and $\eta_{\mathrm m}\!=\! P^{2}/(1-P^{2})$. The upper
and lower limits correspond to the electron tunneling out either conserving
its original spin ($\eta_{\mathrm m}$) or with opposite spin ($-P^{2}$).
Figure 2 shows the distribution for
 three different values of $P$. In the asymmetric case, the shape of the
 distribution is strongly dependent on the ratio $D_{L}/D_{R}$ (not shown).
 
\begin{figure}
\begin{center}
\leavevmode
\epsfxsize = 7.5cm 
%\hsize
\epsffile[31 303 554 650]{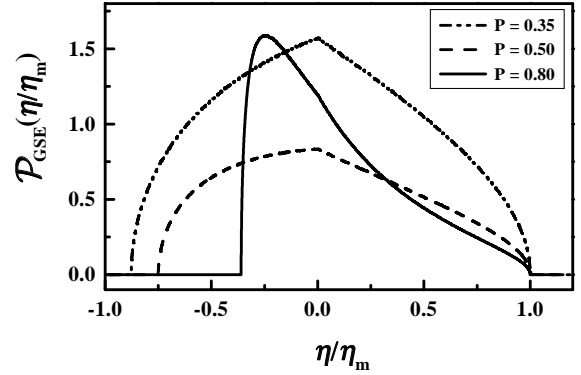}
%\epsfbox{fig3.epsi}
\end{center}
\caption{Probability density of TMR for symmetric barriers and same
ferromagnets with $P  \!=\! 0.35$, $0.5$, $0.8$ in the symplectic ensemble
(GSE). The TMR is scaled  with $\eta_{\rm m}^{} \!=\! P^{2}/(1-P^{2})$. The
distribution for $P \!=\! 0.35$ is multiplied by $2$ for purposes of 
comparison.  The functional form of the distribution depends strongly on
the value of  $P$. In all cases, there is a cusp at $\eta \!=\! 0$. Note
that  $\eta$ can be both positive and negative and significantly different 
from zero. }
\end{figure}

We wish to emphasize three aspects of ${\cal P}_{\mathrm GSE}(\eta)$: (a) The
TMR can be either positive or negative for any polarization as a consequence of
the spin-rotation inside the grain---note that 
${\cal P}_{\mathrm GSE}(\eta\!>\!0)\!=\!{\cal P}_{\mathrm GSE}(\eta\!<\!0)$. (b)
The functional form is strongly affected by SO---note in particular the
disappearance of the divergence at the maximum value and the cusp at zero
(which only vanishes for $P\!\rightarrow \!1$). (c) {\it Even for strong SO
the TMR can be close to its maximum zero-SO value}; this is completly different
from the situation in the classical regime, where the TMR in the presence of
SO is always smaller than in its absence\cite{Brataas2} (note that the GSE
represents the strongest SO coupling). This robustness of the mesoscopic TMR
against SO is a consequence of tunneling through a single level.

{\it Non-Collinear Magnetizations---}
If the magnetizations of the two ferromagnetic leads are not
collinear, a rate equation \emph{does not} apply (even in
the absence of SO) for the same reason as in the SO case---tunneling involves
a coherent superposition of spin states. Instead, we must use Eq.
(\ref{G2}). We assume the magnetization on the left lead points in the
$z$-direction while on the right lead it is tilted by an angle $2 \theta$.
The spin channels on the right lead are then given by
   $\left| \widehat{n},\uparrow \right\rangle \!=\! 
       (\cos \theta,\sin \theta)^{\mathrm T}$, 
   $\left| \widehat{n},\downarrow \right\rangle \!=\! 
       (-\sin \theta,\cos \theta)^{\mathrm T}$. 
Writing $\Psi_{1}({\mathbf r}_{R})$ and $\Psi_{2}({\mathbf r}_{R})$
in this basis, e.g. $\Psi_{1}({\mathbf r}_{R}) \!=\! (\phi_{R}\cos
\theta+\chi_{R}\sin \theta )\left| \widehat{n}, \uparrow \right\rangle +
(-\phi_{R}\sin \theta +\chi_{R}\cos \theta)\left| \widehat{n},\downarrow
\right\rangle$, we readily obtain $\Gamma_{R}$ and then the conductance. Using 
the basis where $\chi_L\!=\!0$ we get
\begin{equation}
G_{\mathrm peak}(2\theta)\!=\!G_{\mathrm peak}(2\theta_{\rm m})
\frac{1+(1-P^2)\xi-P^2{\cal B}(2\theta_{\rm m})}
 {1+(1-P^2)\xi 
-P^2{\cal B}(2 \theta)}
\label{Gtheta}
\end{equation}  
with 
${\cal B}(x)\!=\!\cos{x} \cos\theta_{\rm g}+\cos\varphi_{\rm g} \sin{x}
\sin\theta_{\rm g}$. 
Here, $\theta_{\rm g}$ and $ \varphi_{\rm g}$
describe the rotation of the spin inside the grain in going from left to
right. Notice that the maximum is shifted in the
 presence of SO---the maximum (or minimum) conductance occurs for
 $\tan (2 \theta_{\rm m})\!=\! \tan (\theta_{\rm g})\cos (\varphi_{\rm g})$---and 
that it becomes narrowed as $P\!\rightarrow\!1$. It is worth mentioning that Eq. (\ref{Gtheta}) is different from the result obtained in the classical regime\cite{Brataas}. 
Fig. 3 shows $G_{\mathrm peak}(2\theta)$ for different polarizations both in the
absence and in the presence of SO.

This angular dependence might be used to study the spin-structure of \emph{single} wave functions. 
A spin-polarized STM \cite{SP-STM} could 
provide a spin-resolved image of a single wave function on the surface of a 
small grain, carbon nanotube or quantum corral. In principle it might also be 
used for imaging ferromagnetic quantum dots. This potential for revealing the 
spin structure of a single wave function deserves further investigation.
 
We appreciate helpful conversations with I. L. Aleiner, M. Johnson, K. A. Matveev and D. C. Ralph.
GU acknowledges financial support from CONICET (Argentina).
 
\begin{figure}
\begin{center}
\leavevmode
\epsfxsize = 7.5cm 
%\hsize
\epsffile[35 315 565 651]{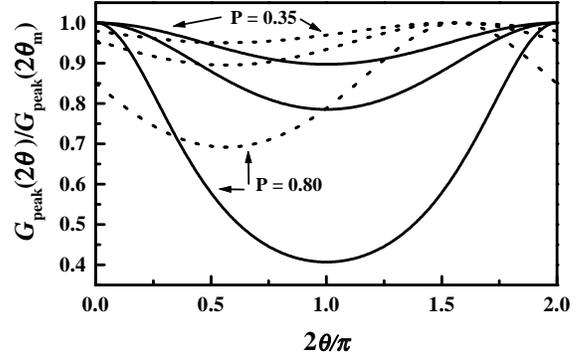}
%\epsfbox{fig4.epsi}
\end{center}
\caption{Conductance as a function of the angle $2\theta$ between 
the magnetization of the leads for different values of $P$. The solid 
(dashed) line shows a typical random realization in the GOE (GSE). The 
maximum occurs for $\tan (2 \theta_{\rm m}) \!=\! \tan (\theta_{\rm g})\cos 
(\varphi_{\rm g})$ where $\theta_{\rm g}$ and $\varphi_{\rm g}$ describe the 
rotation of the spin inside the grain.  Note that $\theta_{\rm g} \!=\! 
\varphi_{\rm g} \!=\! 0$ for the GOE and that the maximum becomes narrowed as 
$P\!\rightarrow\!1$. }
\end{figure}


\begin{thebibliography}{20}

\bibitem{Prinz}
For recent reviews see G. A. Prinz, \emph{Science} \textbf{282}, 1660
(1998); M. Johnson, IEEE Spectrum, \textbf{37(2)}, 33 (2000).
  
\bibitem{Johnson}
M. Johnson, Phys. Rev. Lett. \textbf{70}, 2142 (1993).

\bibitem{Ono}
K. Ono, H. Shimada, S. Kobayashi, and Y. Ootuka, 
   J. Phys. Soc. Jpn. \textbf{65}, 3449 (1996); 
K. Ono, H. Shimada and Y. J. Ootuka, 
   J. Phys. Soc. Jpn. \textbf{66}, 1261 (1997).

\bibitem{Fert}
L. F. Schelp, A. Fert, F. Fettar, P. Holody, S. F. Lee, J. L. Maurice,
   F. Petroff and A. Vaur\`{e}s, Phys. Rev. B \textbf{56}, R5747 (1997).

\bibitem{Ralph}
S. Gu\'{e}ron, M. M. Deshmukh, E. B. Myers and D. C. Ralph, Phys. Rev. Lett.
\textbf{83}, 4148 (1999).
 
\bibitem{Julliere}
M. Julliere, Phys. Lett. \textbf{54} A, 225 (1975); 
J. C. Slonczewski, Phys.  Rev. B \textbf{39}, 6995 (1989).

\bibitem{Takahashi}
S. Takahashi and S. Maekawa, Phys. Rev. Lett. \textbf{80}, 1758 (1998).

\bibitem{Wang}
X. H. Wang and A. Brataas, Phys. Rev. Lett. \textbf{83}, 5138 (1999).

\bibitem{Barnas}
J. Barnas and A. Fert, Phys. Rev. Lett. \textbf{80}, 1058 (1998).

\bibitem{Brataas2}
A. Brataas and X. H. Wang, cond-mat/0004082.

\bibitem{Inamura}
H. Imamura, S. Takahashi and S. Maekawa, Phys. Rev. B \textbf{59}, 6017 (1999).

%\bibitem{BWang}
%B. Wang, J. Wang and H. Guo, cond-mat/9910315

\bibitem{Jalabert}
R. A. Jalabert, A. D. Stone, and Y. Alhassid, 
   Phys. Rev. Lett. \textbf{68}, 3468 (1992).

\bibitem{CBrevs}
For reviews see L. P. Kouwenhoven, {\it et al.}, in 
   {\it Mesoscopic Electron Transport}, edited by L. L. Sohn, 
   L. P. Kouwenhoven, and G. Sch\"on (Kluwer, Dordrecht, 1997) pp. 158-176; and
H. Grabert and M. H. Devoret, \emph{Single Charge Tunneling}
   (Plenum, New York, 1992).

\bibitem{Baranger}
P. W. Brouwer, Y. Oreg and B. I. Halperin, Phys. Rev. B \textbf{60},
R13977 (1999); H. U.  Baranger, D. Ullmo and L. Glazman, Phys. Rev. B \textbf{61}, R2425
(2000); I. L. Kurland, I. L. Aleiner and B. L. Altshuler,
cond-mat/0004205.

\bibitem{SP-STM}
S. Heinze, M. Bode, A. Kubetzka, O. Pietzsch, X. Nie, S. Bl\"{u}gel, 
   and R. Wiesendanger, Science \textbf{288}, 1805 (2000).

\bibitem{halfmetallic}
R. A. de Groot, F. M. Mueller, P. G. van Engen, and K. H. J. Buschow, Phys.
Rev. Lett. \textbf{50}, 2024 (1983)
 
\bibitem{Beenakker}
C. W. J. Beenakker, Phys. Rev. B \textbf{44}, 1646 (1991). 

\bibitem{Glazman}
L. I. Glazman and K. A Matveev, Pis'ma Zh. Eksp. Teor. Fiz. \textbf{48}, 403 (1988) 
[JETP Lett. \textbf{48}, 445 (1988)];
H. Akera, Phys. Rev. B \textbf{59}, 9802 (1999); \textbf{60}, 10683
(1999). 

\bibitem{factor}
This numerical factor takes the Coulomb correlation into
account (no double occupancy). It depends on the change of the spin of the dot;
for instance, $\lambda_{0\rightarrow 1/2} \!=\! (3-2\sqrt{2})$ and
$\lambda_{1/2\rightarrow 1} \!=\! (15/2-3\sqrt{6})$ 

\bibitem{Teresa}
Since $P_q$ is a property of the ``lead + tunnel-barrier'', it can be significantly 
different from the bulk spin-polarization; J. M. De Teresa, A.
Barth\'{e}l\'{e}my, A. Fert, J. P. Contour, R. Lyonnet,
F. Montaigne, P.
Seneor, and A. Vaur\`{e}s, Phys. Rev. Lett \textbf{82}, 4288 (1999).

\bibitem{Meir}
Y. M. Meir and N. S. Wingreen, Phys. Rev. Lett. \textbf{68}, 2512(1992); 
H. Haug and A.-P. Jauho, \emph{Quantum Kinetics in Transport and Optics 
   of Semiconductors} (Springer-Verlag, New York, 1998), pp. 157-178.

\bibitem{Pastawski}
H. M. Pastawski, Phys. Rev. B \textbf{46}, 4053 (1992)

\bibitem{Davies}
J. H. Davies, S. Hershfield, P. Hyldgaard, and J. W. Wilkins,
   Phys. Rev. B \textbf{47}, 4603 (1993).

\bibitem{Brataas}
A. Brataas, Yu. V. Nazarov and G. E. W. Bauer, 
   Phys. Rev. Lett. \textbf{84}, 2481 (2000) and cond-mat/0006174.

\end{thebibliography}
\end{document}